\documentclass[conference]{IEEEtran}
\pdfoutput=1
\IEEEoverridecommandlockouts
\usepackage{cite}
\usepackage{amsmath,amssymb,amsfonts}
\usepackage{algorithmic}
\usepackage{graphicx}
\usepackage{tabularx}
\usepackage{booktabs}
\usepackage[nolist,nohyperlinks]{acronym}
\usepackage{textcomp}
\usepackage[table]{xcolor}
\usepackage{lipsum}
\usepackage{multirow}
\usepackage{listings}
\usepackage[justification=centering,singlelinecheck=false,font=footnotesize]{caption}
\usepackage[inline,shortlabels]{enumitem}

\def\BibTeX{{\rm B\kern-.05em{\sc i\kern-.025em b}\kern-.08em
    T\kern-.1667em\lower.7ex\hbox{E}\kern-.125em}}

\definecolor{highlight}{RGB}{60,153,220}
\definecolor{highlight2}{RGB}{102,211,250}

\newif\ifannotations

\ifannotations
\usepackage{todonotes}
\usepackage{hyperref}
\paperwidth=\dimexpr \paperwidth + 6cm\relax
\oddsidemargin=\dimexpr\oddsidemargin + 3cm\relax
\evensidemargin=\dimexpr\evensidemargin + 3cm\relax
\marginparwidth=\dimexpr \marginparwidth + 3cm\relax
\else
\usepackage[hidelinks]{hyperref}
\usepackage[disable]{todonotes}
\fi

\IEEEaftertitletext{\vspace{-0.5\baselineskip}}
\begin{document}

\bstctlcite{IEEEexample:BSTcontrol} %
	
\begin{acronym}
    \acro{NVD}{National Vulnerability Database}
    \acro{IoT}{Internet-of-Things}
    \acro{CVE}{Common Vulnerabilities and Exposures}
    \acro{CPE}{Common Platform Enumeration}
    \acro{DDoS}{Distributed Denial of Service}
    \acro{FACT}{Firmware Analysis and Comparison Tool}
    \acro{ISA}{Instruction Set Architecture}
    \acro{GPL}{GNU General Public License}
    \acro{SDK}{Software Development Kit}
\end{acronym}

\title{Towards Reliable and Scalable Linux Kernel CVE Attribution in Automated Static Firmware Analyses\\
}

\author{\IEEEauthorblockN{René Helmke and Johannes vom Dorp}
\IEEEauthorblockA{Fraunhofer FKIE, Bonn, Germany, Email: \{rene.helmke, johannes.vom.dorp\}@fkie.fraunhofer.de}
}

\maketitle
\thispagestyle{plain}
\pagestyle{plain}

\begin{abstract}
	In vulnerability assessments, software component-based CVE attribution is a common method to identify possibly vulnerable systems at scale.
    However, such version-centric approaches yield high false-positive rates for binary distributed Linux kernels in firmware images.
    Not filtering included vulnerable components is a reason for unreliable matching, as heterogeneous hardware properties, modularity, and numerous development streams result in a plethora of vendor-customized builds.
    To make a step towards increased result reliability while retaining scalability of the analysis method, we enrich version-based \acs{CVE} matching with kernel-specific build data from binary images using automated static firmware analysis.
    We open source an attribution pipeline that gathers kernel configuration and target architecture to dry build the present kernel version and filter \acsp{CVE} based on affected file references in record descriptions.
    In a case study with 127 router firmware images, we show that in comparison to naive version matching, our approach identifies 68\% of all version \acs{CVE} matches as false-positives and reliably removes them from the result set.
    For 12\% of all matches it provides additional evidence of issue applicability.
    For 19.4\%, our approach does not improve reliability because required file references in \acsp{CVE} are missing.
\end{abstract}

\section{Introduction}
Safety, security, and privacy threats arise alongside embedded system markets.
Growing device numbers\footnote{\scriptsize\url{https://iot-analytics.com/number-connected-iot-devices/}, Accessed: 2022-09-05.} inflate attack surfaces, raising impact and scope of newly found software vulnerabilities in domains pivotal to society~\cite{overview_iot_giri,production_iot_survey_xu,health_care_iot,empirical_study_iot_security_neshenko}.
Thus, it is important to maintain the software security of these systems.

Embedded devices boot into \emph{firmware}: lightweight software packages tailored to the device's use case and limited resources~\cite{firmup}.
Similar to general purpose operating systems, they consolidate services, drivers, and interfaces, but in contrast are highly specialized towards their purpose.
Minimizing costs, vendors commonly rely on \acp{SDK}, which make heavy use of open source.
This gives the advantage of large development communities quickly providing patches when security issues arise.

Linux serves as configurable, modifiable, slim, and powerful system layer in many firmware images~\cite{large_scale_analysis_costin,hrsr2020}.
It is open source, modular, and supports flavors of the MIPS and ARM \acp{ISA} commonly used in embedded devices. %
Vendors configure kernel builds to include all required functions, remove unused components, and add custom functionality.
The compiled kernel is then distributed in binary form as part of the firmware.

As of 2022, there are over 2,900 \ac{CVE} documenting security issues of varying severity affecting different Linux kernel versions~\cite{cve_mitre}.
Yet, the firmware of various widely spread devices contains obsolete or end of life kernels~\cite{large_scale_analysis_costin,hrsr2020}.
This raises questions about device exploitability regarding well-known vulnerabilities.

Intuitively, reproducible exploitation of the target provides undeniable proof for \ac{CVE} attribution.
However, science has yet to find scalable and effective methods for such dynamic verification in binary analysis~\cite{challenges_dynamic_wright,qasem_survey_vuln_detection}.
Heterogeneous system properties require custom solutions:
For example, exploit code may be unavailable for the target \ac{ISA}~\cite{qasem_survey_vuln_detection}.
Resource constraints and missing build chains pose further challenges, as devices might be the only testing platforms available\cite{challenges_dynamic_wright}.
Emulation and re-hosting are options, but require substantial efforts to establish device compatibility~\cite{firmadyne,firmae,challenges_dynamic_wright}.

Static analysis serves heuristics to find \emph{imperfect} proof for \ac{CVE} attribution, e.g., verifying code presence\cite{static_effectiveness_pretschner}.
However, many approaches do not scale well as they require considerable manual work and deep knowledge of each \ac{CVE}~\cite{qasem_survey_vuln_detection}.
Parts are automatable, but needed data may be unavailable or incorrect in repositories~\cite{patch_location_tan,unreliable_cpe}.
Also, automation becomes increasingly challenging in lights of proprietary formats, obfuscation, compiler optimizations, and symbol stripping~\cite{firmup,qasem_survey_vuln_detection}.

In lack of applicable methods, large-scale approaches like firmware analysis tools~\cite{fact,onekey,netrise} and studies~\cite{hrsr2020,large_scale_analysis_costin,large_scale_zhao} commonly attribute bugs by matching versions against \ac{CVE} databases.
Trading in reliability for applicability, their results need manual verification but may raise awareness for potential security issues.

In 2020, we used version matching as part of a large-scale empirical study on home router security~\cite{hrsr2020}.
For the Linux kernel, this method proves exceptionally unreliable:
Due to custom build configurations, we can not assume that kernels include components flagged as vulnerable for the subject version.
E.g., \acp{CVE} affecting hardware drivers that vendors removed from their builds are not applicable, but the method does not cover this factor -- leading to false-positive matches.

Qualitative evaluation of our approach showed high false-positive rates as build customizations for a control group of firmware samples were taken into account.
While the large result set might still be useful for analysts, more precise methods are needed to reduce manual verification efforts~\cite{qasem_survey_vuln_detection}.

To improve the reliability of version-based Linux \ac{CVE} attribution in large-scale scenarios, we enrich the process with kernel-specific data from automated static firmware analysis.
We extract kernel configurations from binary images and reconstruct the kernel build process to identify included components with file-level granularity.
Hereby, we reduce the set of false-positive matches requiring further manual verification.

In the following, we provide:
\begin{enumerate}
    \item A description of our methodology for Linux \ac{CVE} attribution, based solely on binary kernel representations.
    \item An empirical evaluation in which we compare our approach against naive version-based \ac{CVE} matching using the 2020 Home Router Security Report~\cite{hrsr2020} dataset.
\end{enumerate}
Furthermore, we release the implementation of the methodology used for the evaluation as open source to the public\footnote{\scriptsize\url{https://github.com/fkie-cad/cve-attribution-s2}, Accessed: 2022-09-05.}.

\section{Background \& Related Work}
\label{sec:related_work}

There are various automation approaches to discover and attribute security issues in binary targets.
Aside of version-based matching against vulnerability repositories, they include code similarity and patch analysis~\cite{code_similarity_survey}, taint analysis and symbolic execution~\cite{symbolic_execution_survey}, fuzzing~\cite{fuzzing_survey_manes}, and various emulation-based methods~\cite{challenges_dynamic_wright}.
While we acknowledge substantial work done in these directions, we point out their limited applicability in large-scale analyses due to resource constraints, low emulation success rates, and missing bootstrap data~\cite{qasem_survey_vuln_detection,challenges_dynamic_wright}.

Focusing on large-scale attribution of known security issues in firmware, this section first discusses the scientific need for sound data as foundation of reliable bug attribution~(\ref{sec:related_work:sound_data}).
Then, we present attribution methods applied in large-scale firmware security studies~(\ref{sec:related_work:largescale}).

For comprehensive surveys on state of the art research and the problem space, especially in terms of scalability and automation, we refer to Qasem et al.~\cite{qasem_survey_vuln_detection} and Wright et al.~\cite{challenges_dynamic_wright}.

\subsection{Sound Data as Foundation for Reliable Bug Attribution}
\label{sec:related_work:sound_data}
Detail information on known vulnerabilities lays the foundation of reliable bug attribution~\cite{qasem_survey_vuln_detection}.
The community-driven \ac{CVE} catalog~\cite{cve_mitre} is the largest data source and de facto standard for vulnerability identification.
For each tracked issue, there is an identifier, description, and follow-up references.
The derivative \ac{NVD}~\cite{nvd} adds data on severity and impact scores, but also lists affected products and their versions.
The latter is encoded in an identification scheme, the \ac{CPE}~\cite{nvd}.

Records vary in information quantity and quality.
E.g., Sanguino and Uetz~\cite{unreliable_cpe} show that errors in \ac{CPE} assignments, but also data inconsistencies harm soundness.
Code similarity approaches like~\cite{firmup} exemplify that pivotal data can be missing or is hard to obtain, as they require access to ground truth embedded in commits and patches.
If \acp{CVE} affect closed source projects, issuers will not share technical details on fixes in public.
For open source, Tan et. al.~\cite{patch_location_tan} show that code collection remains challenging:
Patches are unreferenced, not marked, incorrect, or hide in bug tracker discussions.

Thus, researchers construct small datasets of selected \acp{CVE} that their proposed techniques can ingest for evaluation~\cite{qasem_survey_vuln_detection}.
For each \ac{CVE}, this implies in-depth investigation, additional data aggregation, and technical bug knowledge, which limits the applicability of previously mentioned approaches for large-scale scenarios.
The Vulncode-DB~\cite{vulncode_db} addressed this issue and annotated \ac{CVE} data with fine-granular technical information, but got discontinued due to bootstrapping problems.

As for Linux kernel \acp{CVE}, issuers embed references to bug locations in the project's source tree in \ac{CVE} short descriptions~\cite{cve_mitre}.
In combination with extracted build configurations from firmware, the proposal in this paper takes advantage of this observation to identify components built into the kernel.

The \emph{linuxkernelcves.org}~\cite{linuxkernelcves} project argues that for Linux's many version streams, contributors, and distributors, there is inaccuracy in \ac{CPE} data to precisely track affected versions.
They crawl commit references in \acp{CVE} from Linux distribution sources, e.g., Ubuntu and Android bug trackers, to determine the first patch appearance in mainline kernels.
However, while~\cite{linuxkernelcves} host a regularly updated dataset, they stress that their \ac{CVE} post-processing code has neither been evaluated nor published yet -- result reliability is unknown.

\begin{figure*}[t!]
    \centering
    \includegraphics[width=1.0\textwidth]{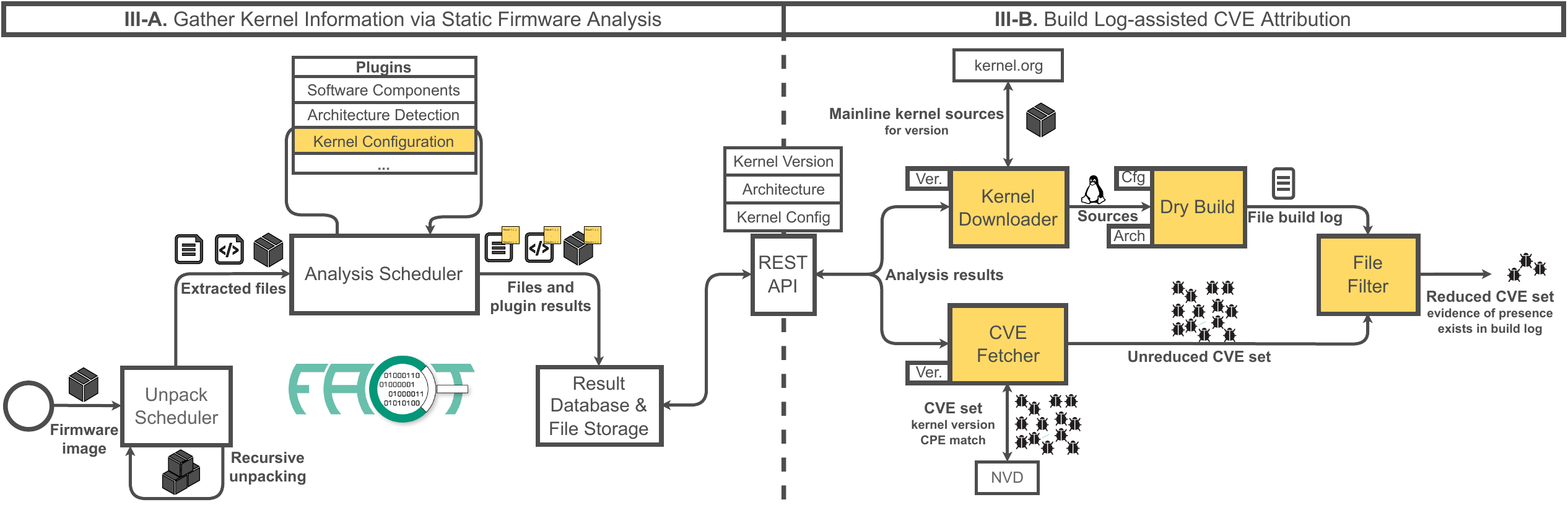}
    \caption{Our methodology describes a two-staged static analysis pipeline: First, we extract, analyze, and annotate ingress firmware images to detect kernel version, configuration, and architecture (left). Then, we fetch the corresponding mainline kernel sources, install the configuration, and perform a dry build to generate a file build log. In combination with a version-filtered list of potentially attributable \acp{CVE}, we eliminate all records that do not include affected component source files we witnessed during the dry build.}
    \label{fig:1:methodology_overview}
\end{figure*}

\subsection{Static Bug Attribution in Large-Scale Firmware Analyses}
\label{sec:related_work:largescale}

In 2014, Costin et al.~\cite{large_scale_analysis_costin} execute a quantitative study on embedded device security by analyzing 32.000 firmware images.
They implement pattern-based heuristics and specify indicators of security malpractice in firmware images, e.g., hardcoded passwords, cryptographic material, but also included application version numbers to attribute \acp{CVE}.
Their static analysis yields \ac{CVE} matches in userspace applications of 693 firmware images.
While they inspect Linux kernel versions, these are not included in the bug attribution.
Furthermore,~\cite{large_scale_analysis_costin} report unsolved challenges in result verification, as not only \ac{CVE} data is incomplete, but also vendors may custom-patch affected files.
Since 2014, cross-architecture code similarity methods have drastically improved and may be used as imprecise measure for verification in this case, e.g., FirmUp~\cite{firmup}.
However, acquiring and processing patches for thousands of \acp{CVE} to bootstrap code similarity methods deems infeasible based on the imprecise \ac{CVE} repositories.

Zhao et al.~\cite{large_scale_zhao} develop FirmSec, a large-scale static analysis pipeline to empirically study vulnerabilities introduced by third-party components and software in \acs{IoT} devices.
Here, a novelty is that they identify software versions by extracting syntactical and control-flow graph features.
They construct a repository of 1,191 third party components and then extract matching features out of the component's corresponding source files.
However, they neither consider vendor-specific build configurations nor the Linux kernel.

Similar to~\cite{large_scale_analysis_costin}, we assess and compare the state of firmware security of 127 home routers available on German markets in our whitepaper Home Router Security Report 2020~\cite{hrsr2020}.
Using the automated~\ac{FACT}~\cite{fact}, we also implement pattern-based heuristics to detect best practice violations in firmware images.
E.g., we analyze the usage of exploit mitigation techniques like stack canaries and non-executable bits in userspace binaries.
To quantize the differences between used Linux kernels in devices from a security perspective, we identified their versions, matched them against the \ac{NVD}~\cite{nvd} using \ac{CPE}, and reported the results.
Subsequent discussions and observations proved that attributing CVEs only based on kernel version leads to high false-positive rates and in effect to high manual verification effort and low chance of adoption.
On top of the issues of uncertainty in \ac{CVE} databases and unidentifiable vendor patches, we found that the kernel's high modularity, heterogeneity, and vendor-specific builds inflated false-positives:
Qualitative analysis showed that we matched for \acp{CVE} in components that were verifiable absent in the build -- as for some firmware samples we were able to extract the kernel configuration to identify included components. %

We could not find any related work examining this Linux kernel-specific problem in \ac{CVE} attribution.
Yet, we observed consistent information in databases to identify components and their files affected by a \ac{CVE}.
In combination with our observation of extractable build configurations, we identify a research gap and possibility to reliably reduce false-positive matches in kernel \ac{CVE} attribution.

\section{Methodology}
\label{sec:methodology}

This section describes our proposed methodology to enrich the version-based Linux kernel \ac{CVE} attribution process with build-specific annotations.
We show an automated static analysis pipeline that finds and extracts kernel configurations, dry builds the found kernel version, and filters \acp{CVE} based on affected version and build log-included files.

Figure~\ref{fig:1:methodology_overview} provides an overview of our methodology.
We establish a two-stage process:
In the first and left-hand stage, we unpack, analyze, and annotate each file of an ingress firmware image.
Gathered information includes Linux kernel version, \ac{ISA}, and kernel build configuration.
In the second and right-hand stage, we leverage upon said data to perform the actual \ac{CVE} attribution and filtering step.
Yellow boxes in Figure~\ref{fig:1:methodology_overview} mark components this paper contributes.

In the two following subsections~\ref{sec:methodology:static_analysis} and~\ref{sec:methodology:attribution}, we provide
\begin{lstlisting}[frame=single,caption={Software Components YARA rule to identify Linux kernel versions. \emph{no\_text\_file} is a custom rule excluding \emph{text/*} MIME types.},captionpos=b,basicstyle={\ttfamily \scriptsize},breaklines=true,numbers=left, xleftmargin=7mm,framexleftmargin=6mm]
strings:
 $v = /Linux version \d\.\d{1,2}\.\d{1,3}(-[\d\w.-]+)?/ nocase ascii wide
    
condition:
 $v and no_text_file
\end{lstlisting}
detailed technical insights on each stage and step.
We finish with a short example of CVE-2017-17864 in~\ref{sec:methodology:example} to demonstrate the added attribution reliability our approach offers.

\subsection{Gather Kernel Information via Static Firmware Analysis}
\label{sec:methodology:static_analysis}

For stage one, we apply and enhance the firmware analysis tool \ac{FACT}~\cite{fact}, which is maintained at our research group.
FACT is open source, comes with a variety of firmware container unpackers, and provides a plugin system with ready-to-use analysis modules we can leverage for kernel version and \ac{ISA} detection.
In the following, we describe all pipeline steps that are of importance for the proposed attribution methodology.

Consider an arbitrary firmware image.
After submitting it to the pipeline, \ac{FACT}'s \textbf{Unpack Scheduler} tries to identify container formats.
Choosing from a palette of custom extractors and integrated tools like binwalk~\cite{binwalk}, it recursively extracts or decompresses file contents until all options are exhausted.

The \textbf{Analysis Scheduler} receives each extracted file and applies a preconfigured set of analysis plugins.
These implement heuristics, e.g., based on YARA~\cite{yara} rules, third party tools like checksec~\cite{checksec}, or custom analysis implementations.
Analyzed files and their annotated results are stored in the \textbf{Result Database \& File Storage}.

Our methodology depends on the results of three plugins: Software Components, Kernel Configuration, and Architecture Detection.
While \ac{FACT} already provides the former and latter, this paper contributes the new Kernel Configuration plugin.

The \textbf{Software Components} plugin applies YARA rules to detect various software components along with their version.
One of the rules, a part of which is shown in Listing 1, detects the Linux kernel and its version.

We contribute the \textbf{Kernel Configuration} plugin, which detects and extracts Linux kernel build information.
Stage two of our pipeline requires this data to filter out \acp{CVE} applicable to components excluded from the build.

\begin{figure}[t!]
    \centering
    \includegraphics[width=1.0\linewidth]{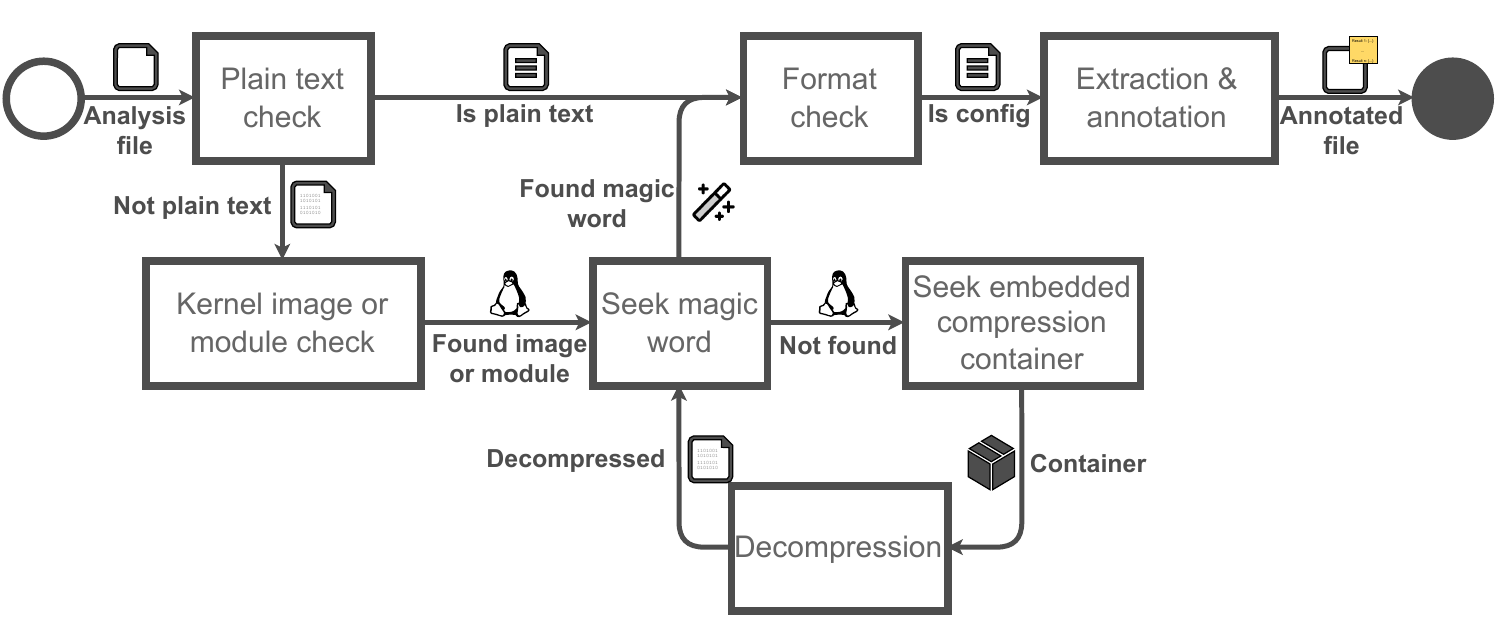}
    \caption{High-level control flow of the Kernel Configuration analysis plugin. It detects configurations in plain text files, kernel images, and modules. Decompression takes place if needed. Results are annotated to the file.}
    \label{fig:2:methodology_ikconfig}
\end{figure}

Figure~\ref{fig:2:methodology_ikconfig} illustrates the plugin's high-level control flow.
Not explicitly stated edge transitions lead to analysis termination.
Inside a firmware, kernel configurations may be present in different plain text and binary formats.
Thus, depending on the file's MIME type, the plugin differentiates between kernel-related binary files and plain text.
Here, plain text is the most straightforward case, as the plugin then checks for headers and build configuration directives.
An example plain text configuration can be found in the Debian Repositories\footnote{\scriptsize\url{https://salsa.debian.org/kernel-team/linux/-/blob/master/debian/config/arm64/config}, Accessed: 2022-09-05.}.

Another way to obtain the configuration is by making use of an enabled {\ttfamily CONFIG\_IKCONFIG}\footnote{\scriptsize\url{https://github.com/torvalds/linux/blob/master/init/Kconfig}, Accessed: 2022-09-05.} directive:
When it is set to {\ttfamily Y} during the firmware kernel's build time, a copy of the configuration is embedded into the binary image -- either as inline string or compressed container.
If it is set to {\ttfamily M}, said information is outsourced to a kernel module.
Thus, if the file is either a kernel image or module, our plugin searches for a magic word that is prepended to the inline string.
Once found, we apply a set of signatures to detect the correct format of the embedded configuration and annotate our findings to the analyzed file.
Otherwise, we seek for embedded containers, try to extract them, and again check for magic word and format.

The \textbf{Architecture Detection} plugin identifies target \acp{ISA} using four different measures.
For executable files in ELF format, it parses the {\ttfamily e\_machine} and {\ttfamily e\_flags} fields of the file header.
For kernel configurations, it searches for the presence of architecture-specific feature directives.
E.g., {\ttfamily ARM64\_USE\_LSE\_ATOMICS} is only supported by the ARMv8 specification.
Furthermore, the plugin detects and parses CPU information from device trees~\cite{devicetree} -- a widely spread bootloader-to-operating-system interface which also lists system hardware and its properties.
Finally, the plugin does also rely on MIME databases to identify any file type that might leak target platform information.

\subsection{Build Log-assisted \ac{CVE} Attribution}
\label{sec:methodology:attribution}

The build log-assisted \ac{CVE} attribution is the second stage of our proposed analysis pipeline in Figure~\ref{fig:1:methodology_overview}.
Here, we first use~\ac{FACT}'s \textbf{REST API} to consolidate the kernel version, kernel build configuration, and detected target \ac{ISA}.

Then, the \textbf{Kernel Downloader} fetches version sources from \emph{kernel.org}.
Note that the assumption of vendors using a mainline, i.e., unmodified, kernel is false in general due to custom patching and additional proprietary components.
Large parts of Linux are licensed under the \ac{GPL} and, thus, modifications must be published\footnote{\scriptsize\url{https://www.gnu.org/licenses/gpl-faq.html}, Accessed: 2022-09-05.}.
However, for two reasons we could not find a scalable way to obtain these versions:
First, some vendors distribute packages through individual processes on request\footnote{\scriptsize\url{https://www.zyxel.com/form/gpl_oss_software_notice.shtml}, Accessed: 2022-09-05.}.
Second, the required build tools do not have to be part of the package -- which is pivotal to our pipeline.

\textbf{Dry Build} is the next step in Figure~\ref{fig:1:methodology_overview}:
We set the target \ac{ISA} and install the extracted kernel build configuration in the downloaded kernel source code.
Then, we execute a dry run via {\ttfamily make}, which does not compile the kernel but prints each compilation recipe instead.
This approach has the advantage of low computational overhead and does not require cross-compilation environment bootstrapping.
The goal of this step is to gather a list of source files from the build log, which our pipeline \emph{witnessed} to be included in the kernel.

The \textbf{CVE Fetcher} executes simultaneously.
Here, we query the \ac{NVD}~\cite{nvd} dataset for all Linux kernel \acp{CVE} and filter out all records that do not refer \acp{CPE} stating the extracted Linux kernel version to be vulnerable.
The resulting set of potentially applicable \acp{CVE} is the output of version-based attribution.

Finally, the \textbf{File Filter} step combines the outputs of CVE Fetcher and Dry Build:
As we observed that Linux kernel \acp{CVE} state the affected source files in their short description, records from the version-based result set can be eliminated.
We reduce it by removing every \ac{CVE} that does not reference an affected file we witnessed in the build log.

\subsection{Short Example - CVE-2017-17863}
\label{sec:methodology:example}

To demonstrate the practical addition of our Linux kernel \ac{CVE} attribution pipeline, we shortly present an example using CVE-2017-17863\footnote{\scriptsize\url{https://nvd.nist.gov/vuln/detail/CVE-2017-17863}, Accessed: 2022-09-05.}.
The record description states an invalid memory access due to an integer overflow in {\ttfamily kernel/bpf/verifier.c}.
Also, \ac{CPE} states that kernel versions from 4.9 to 4.9.71 are vulnerable.

Consider a firmware sample with Linux kernel 4.9.60 and the underlying kernel configuration.
Naive version-based matching reports a positive match as the version is within \ac{CPE} range.
However, if the configuration shows that BPF support is disabled because {\ttfamily CONFIG\_BPF} is set to {\ttfamily N}, all associated source files, including {\ttfamily kernel/bpf/verifier.c}, are excluded from the compilation.
Thus, our proposed attribution method does not witness the affected file presence during the dry build step:
CVE-2017-1786 is a false-positive match and we can eliminate it from the result set.

\section{Case Study}
\label{sec:case_study}

We perform a case study to evaluate the reliability of our enriched version-based Linux kernel \ac{CVE} attribution in large-scale static analyses.
For this purpose, we let our pipeline analyze the Home Router Security Report 2020~\cite{hrsr2020} corpus, which vendors reported to yield high false-positive rates using version-based \ac{CVE} matching.
We raise two research questions:
\begin{itemize}
    \item[\textbf{R1}]Our methodology requires access to specific information in firmware samples and \ac{CVE} repositories.
How many samples and \acp{CVE} fulfill these modalities? \emph{How applicable is our approach in a real-world scenario?}
    \item[\textbf{R2}] With version-based \ac{CVE} matching as baseline, \emph{what impact has the methodology on result reliability?}
\end{itemize}

In the following subsections, we first provide detailed information on our experiment and used dataset~(\ref{sec:case_study:experiment_and_dataset}).
Then, we present the results and analyze them within the context of both stated research questions~(\ref{sec:case_study:rq1}~and~\ref{sec:case_study:rq2}, respectively).

\subsection{Experiment \& Firmware Corpus}
\label{sec:case_study:experiment_and_dataset}
\textbf{Experiment Execution.} We deploy our static analysis \ac{CVE} attribution pipeline on a system with AMD Ryzen 7 2700x processor and 32 GiB DDR4 RAM, running Ubuntu 20.04.4 LTS.
\ac{FACT} v4.0 (commit \texttt{38df4883}) is used in the first pipeline stage.
We extract each firmware of the Home Router Security Report 2020~\cite{hrsr2020} corpus and apply the required analysis plugins CPU Architecture, Software Components, and Kernel Configuration~(cf.,~Section~\ref{sec:methodology:static_analysis}).
The second pipeline stage executes on the same machine based on a snapshot of the~\ac{NVD}~\cite{nvd} -- taken on 2022-08-30.
The snapshot has records for 2,910 Linux kernel~\acp{CVE} attributable through \ac{CPE}.
For each component in our system, we collect details on ingress and egress data, including plugin results, version-based \ac{CVE} matches, and filtering decisions.

\textbf{Firmware Corpus.} The analyzed Home Router Security Report~\cite{hrsr2020} corpus is publicly documented\footnote{\scriptsize\url{https://github.com/fkie-cad/embedded-evaluation-corpus/blob/master/2020/FKIE-HRS-2020.md}, Accessed: 2022-09-05.}, reconstructable, and consists of firmware from 127 home routers available in the european market.
Devices of seven well-known vendors are included: ASUS, AVM, D-Link, Linksys, Netgear, TP-Link, and Zyxel.
Samples were scraped on 2020-03-27.

\begin{figure}[t!]
    \centering
    \includegraphics[width=1.0\linewidth]{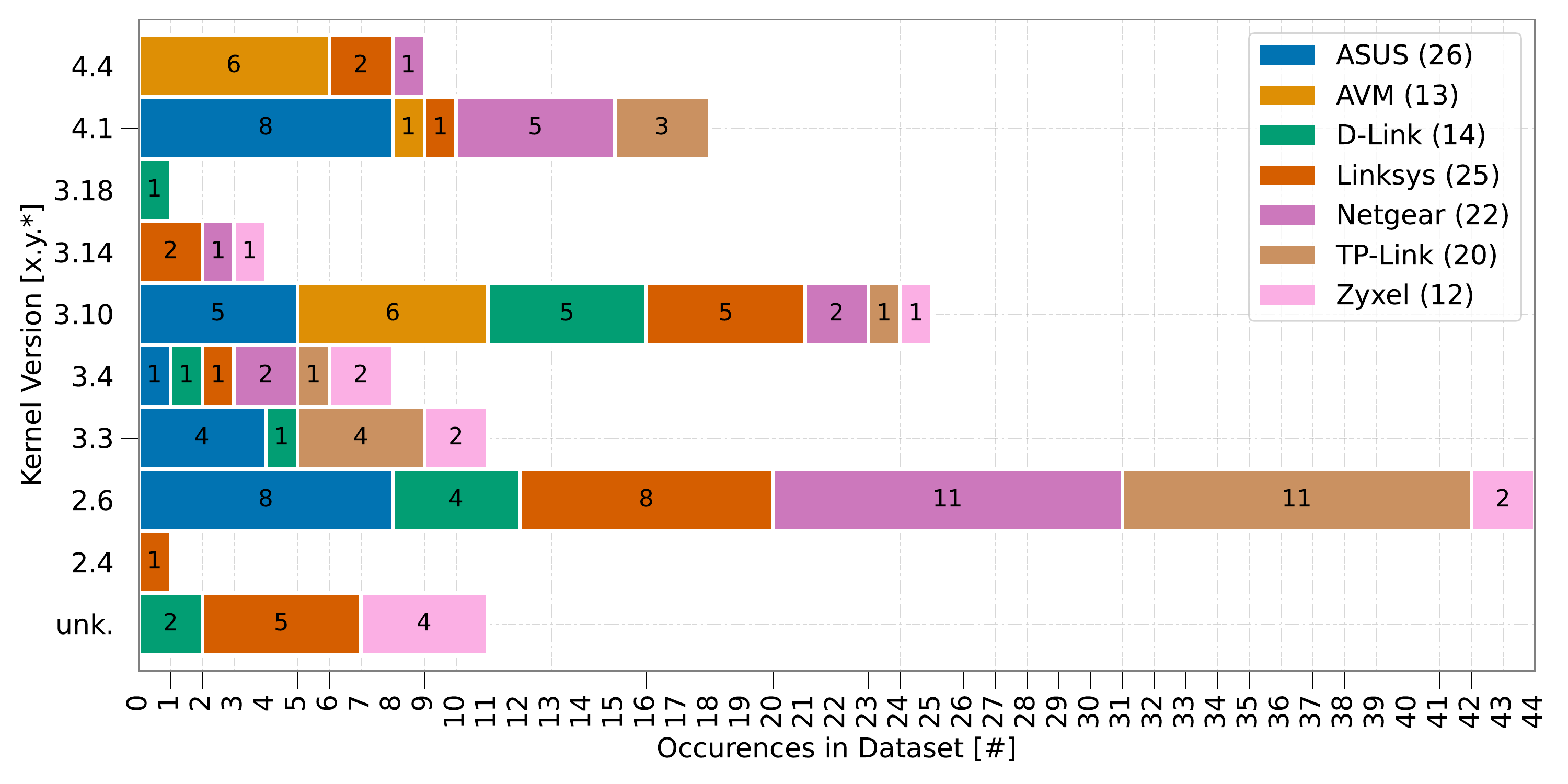}
    \caption{Kernel version distribution of the subject Home Router Security Report 2020 dataset~\cite{hrsr2020}, which was scraped on 2020-03-27. Patch versions are clustered by major and minor release. Across 127 firmware images from 7 different vendors, \ac{FACT} found 121 Linux kernels between v2.4.20 and v4.4.60. For 11 samples, no operating system could be identified (\emph{unk.}).}
    \label{fig:3:linux_versions}
\end{figure}

Figure~\ref{fig:3:linux_versions} shows the distribution of Linux kernel versions detected by \ac{FACT} across all firmware images.
Numbers in parentheses and on top of bars count sample sizes per vendor and kernel.
Patch levels are clustered by major and minor release for illustrative purposes, but are considered during \ac{CVE} attribution.
Across all 127 samples, 121 binary distributed Linux kernels from v2.4.20 to v4.4.60 are included.
11 firmware images are not analyzable due to failed operating system analysis or unpacking errors.
Note that firmware can contain multiple kernels, e.g., embedded devices may consist of subcomponents running their own systems.

\begin{figure}[t!]
    \centering
    \includegraphics[width=1.0\linewidth]{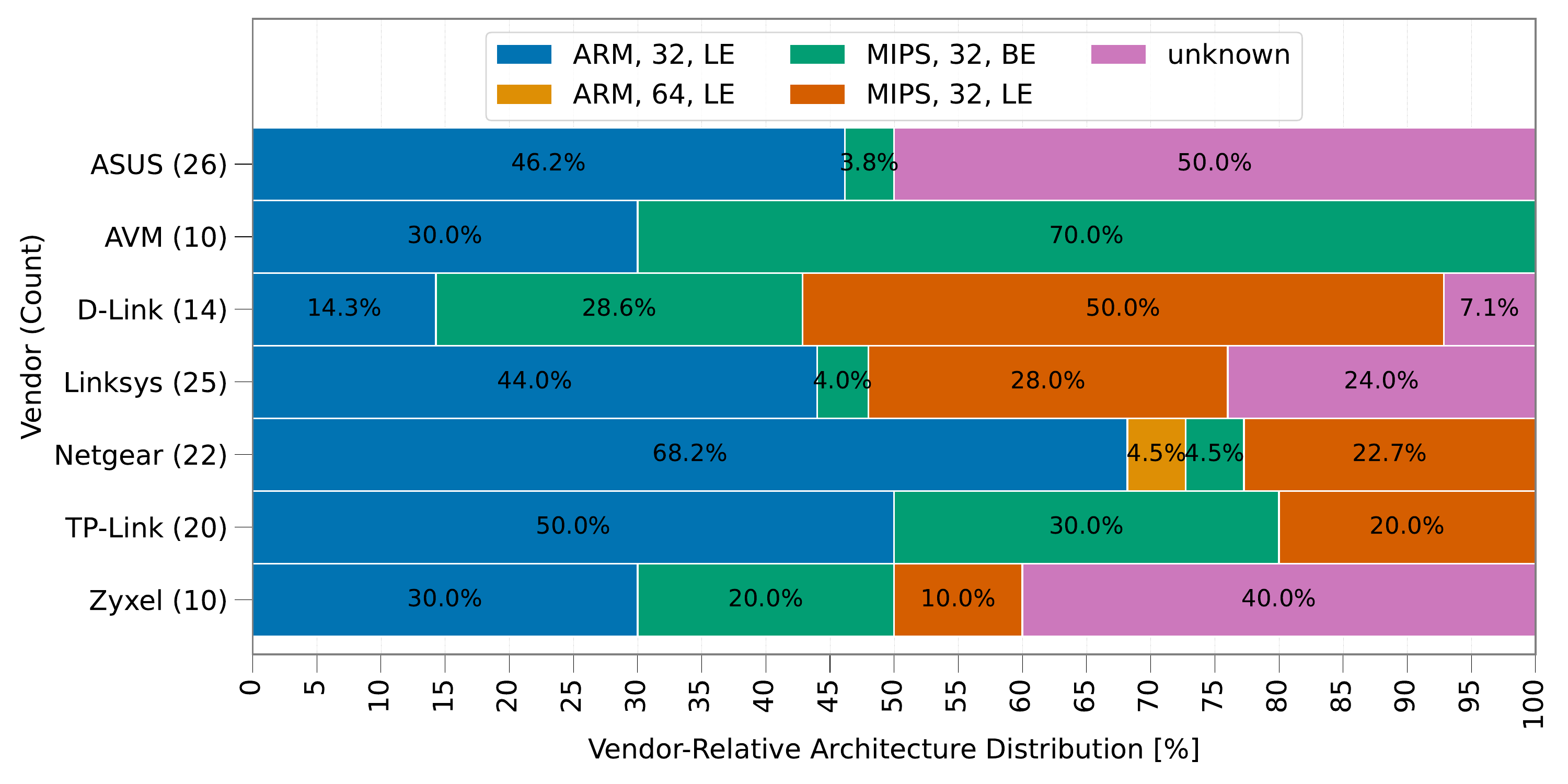}
    \caption{Relative CPU Architecture Distribution across all firmware samples, clustered by vendor. MIPS and ARM, both in big and little endianess are prevalent in this firmware corpus of home router devices.}
    \label{fig:4:architectures}
\end{figure}

Figure~\ref{fig:4:architectures} shows the relative architecture distribution across all samples, clustered by vendor.
MIPS and ARM \acp{ISA} dominate the corpus, both in big and little endian.
It includes different word lengths, but the majority of devices uses 32-bit.
The \ac{ISA} is unknown for 24 samples.
More detailed insights into the corpus can be found in~\cite{hrsr2020}.

We argue that the corpus is of sufficient size to demonstrate applicability, but also choose it to investigate the reliability of matches we reported in~\cite{hrsr2020} using naive version-based Linux kernel \ac{CVE} attribution.
Furthermore, it offers heterogeneous properties beneficial for our case study, as it covers Linux kernels from three major releases, widely-spread \acp{ISA}, and devices of multiple vendors.
A drawback is that other device classes than home routers are missing.

\begin{figure*}[t!]
    \centering
    \includegraphics[width=1.0\textwidth]{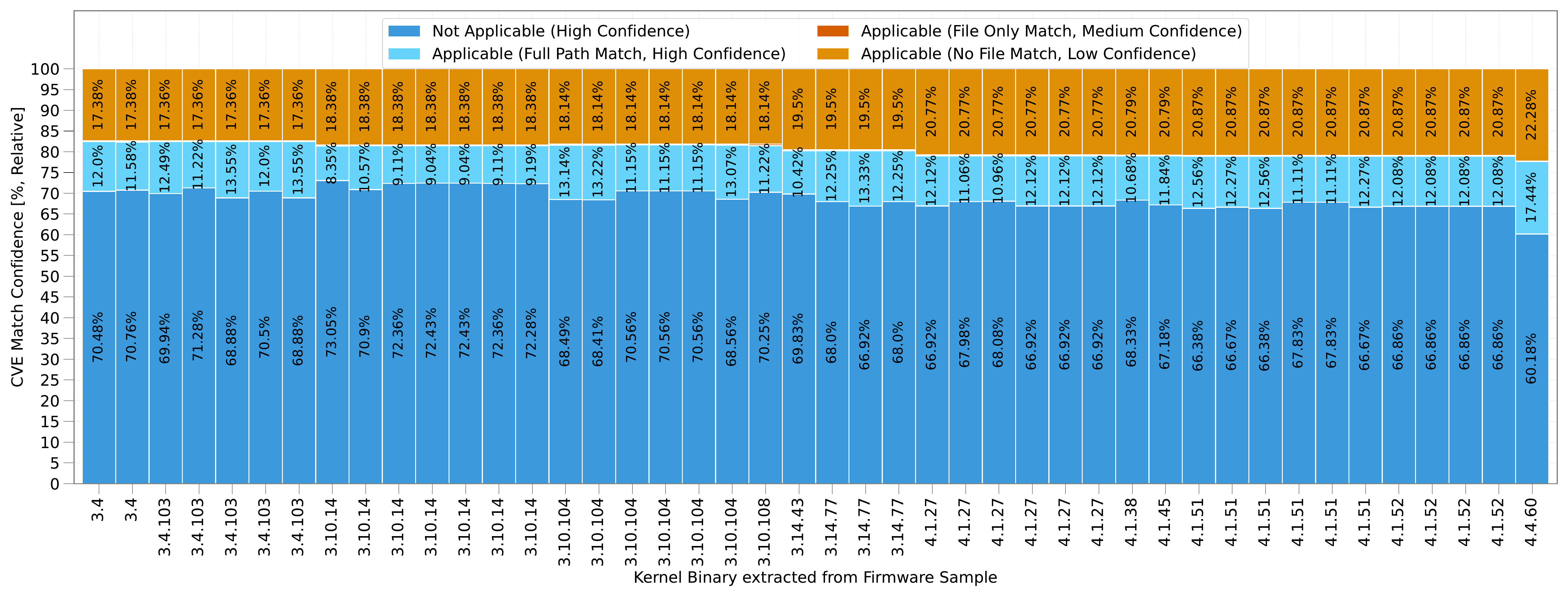}
    \caption{Filter verdict distribution of our pipeline relative to the baseline CVE attribution results for each of the 44 analyzed kernels. Each entry on the horizontal axis represents a unique firmware. We classify them into four different verdict confidence categories. With high confidence, our methodology marks around $68.37\%$ of all baseline matches as false-positive and $12.04\%$ as true-positive candidates (medians). Relative path matches with medium confidence match verdict are negligible with $0.19\%$.}
    \label{fig:5:cve_distribution}
\end{figure*}

\subsection{R1 Analysis -- Applicability in Real-World Scenarios}
\label{sec:case_study:rq1}
How applicable is our approach in real-world large-scale automated firmware analyses?
To find answers to this question we identify two methodological requirements that must be fulfilled for each firmware and Linux \ac{CVE}, respectively:

\begin{enumerate}
    \item[\textbf{S1}] \ac{FACT} firmware extraction and plugin analyses of stage one must succeed to consolidate the kernel version, \ac{ISA}, and kernel configuration.
    \item[\textbf{S2}] \ac{CVE} descriptions must reference affected files to filter vulnerable components not included in the kernel build.
\end{enumerate}

Using the firmware corpus, we evaluate egress and ingress data of each step in the proposed pipeline with regards to these requirements.
Table~\ref{tbl:1:applicability} shows the results.
Highlighted rows designate effective requirement fulfillment rates over all analyzed firmware images and Linux kernel \acp{CVE}.

\begin{table}[t!]
    \centering
        \begin{tabular}{lcc}
            \toprule
            \multicolumn{3}{c}{\textbf{S1 Requirement (FACT Analysis Success)}}\\ \toprule
              & FW Matches [\#] & Fulfilled$\left[\frac{\text{FW Matches}}{\text{FWs Total}}\right]$\\ \midrule
             Extraction & 116 & $0.9133$\\
             Kernel Version & 116 & $0.9133$\\
             Architecture Detection & 103 & $0.8110$\\
             \cellcolor{highlight2}Kernel Configuration & \cellcolor{highlight2}44 & \cellcolor{highlight2}$0.3464$\\\toprule
            \multicolumn{3}{c}{\textbf{S2 Requirement (File Reference in Linux Kernel CVE)}}\\ \toprule
            & \ac{CVE} Matches & Fulfilled$\left[\frac{\text{CVE Matches}}{\text{CVEs Total}}\right]$\\ \midrule
            \cellcolor{highlight2}Full Path Reference & \cellcolor{highlight2}1743 & \cellcolor{highlight2}$0.5990$\\
            \cellcolor{highlight2}File Only Reference & \cellcolor{highlight2}129 & \cellcolor{highlight2}$0.0443$\\
            No Reference & 1038 & $0.3567$\\ \bottomrule
        \end{tabular}
    \caption{Method Applicability Analysis for the Firmware Corpus}
    \label{tbl:1:applicability}
\end{table}

For requirement S1, \ac{FACT} successfully extracts 116 out of 127 firmware images.
The analyses detect at least one kernel version in all extractable firmware, and \acp{ISA} in 103 of them.
However, our Kernel Configuration plugin finds build information in only 44.
As this plugin depends on results from previous analyses~(cf.,~Section~\ref{sec:methodology}), the matches are a subset of the extracted firmware images with identified \ac{ISA} and kernel version.
With missing Kernel Configurations as limiting factor of method applicability, $34.64\%$ of all analyzed firmware samples fulfill requirement S1.
This rate is explainable considering that a) \texttt{IKCONFIG} must be explicitly enabled to embed kernel configurations into binary representations and b) it is common practice for vendors to strip and obfuscate valuable analysis information from firmware.

For requirement S2 (affected files must be referenced in Linux kernel \ac{CVE} descriptions), data analysis over all Linux \acp{CVE} inside the \ac{NVD} yields three different categories:
Files are either referenced as \textbf{Full Path} relative to the kernel's source tree, or the reference is \textbf{File Only} (location in the source tree is unknown), or \textbf{No Reference} exists at all.
Table~\ref{tbl:1:applicability} distributes all 2,910 Linux kernel \acp{CVE} across these classes, showing that the proposed approach is applicable to $1,872$ ($64.33\%$) kernel \acp{CVE}.
For \acp{CVE} with no included file reference, the approach falls back to version-based \ac{CVE} matching and, thus, can not add value to result reliability.

\subsection{R2 Analysis -- Impact on \ac{CVE} Attribution Result Reliability}
\label{sec:case_study:rq2}

With version-based Linux kernel \ac{CVE} matching as baseline, what impact has our attribution pipeline on result reliability?
We approach research question R2 by analyzing the attribution results of all 44 firmware images our methodology is applicable to~(cf.,~Section~\ref{sec:case_study:rq1}).
Subject samples include kernels ranging from v3.4.0 to v4.4.60.
Out of these, only one still receives mainline updates at the time of this evaluation (4.4.x).

The baseline method attributes a median of 1,196 \acp{CVE} per firmware image, which is roughly $40\%$ of \emph{all} Linux kernel \acp{CVE} present in the \ac{NVD}.
A possible explanation lies within unsound and/or unmaintained \ac{CVE} records in the \ac{NVD}~\cite{unreliable_cpe,patch_location_tan}.
Yet, such imprecise data would also imply the existence of false-negatives that should have been attributed to the kernel version, but were not due to incorrect \ac{CPE}.
An argument in favor of these results is the amount of end of life kernels in the dataset.
However, based on the results we present in the following paragraphs, there is reason to assume that the baseline yields exceedingly high false-positive rates.

Version-based \ac{CVE} attribution is an intermediate result of our methodology~(cf.,~Section~\ref{sec:methodology}).
To estimate the impact our pipeline has on result reliability,
we consolidate all decisions of the build-log assisted filtering to classify them into four categories of verdict confidence:

\begin{itemize}
    \item \textbf{Applicable (High)} -- \ac{CVE} references affected files and full file path is witnessed in build log.
    \item \textbf{Not Applicable (High)} -- \ac{CVE} references affected files but none of them is present in the build log.
    \item \textbf{Applicable (Medium)} -- \ac{CVE} references affected files, but does not state full file paths. A file was matched and seen in the build log, but ambiguity exists due to duplicate names in the source tree.
    \item \textbf{Applicable (Low)} -- No file references, we can not decide on applicability and fall back to version-based matching.
\end{itemize}
The idea is to map persuasiveness of additional evidence the pipeline gathers within a trial:
File matches are witnesses for \ac{CVE} applicability, but not every match is equally credible.

Figure~\ref{fig:5:cve_distribution} shows the filter verdict distribution of our pipeline relative to the baseline \ac{CVE} attribution results for each analyzed kernel.
Versions are ordered from oldest (left) to newest (right).
Note that a single kernel was found in each one of the 44 analyzed samples.
Thus, each entry on the horizontal axis represents a unique firmware.
In the following, all discussed distribution values are medians across all samples.

The proposed Linux kernel \ac{CVE} attribution methodology is able to make a medium to high confidence decision for $80.6\%$ of all version-based matches.
The portion of high confidence applicable \ac{CVE} matches is $12.04\%$.
Relative path matches yielding medium confidence applicability are negligible with $0.19\%$.
As indicated by the bottom bars belonging to the class of Not Applicable (High), our pipeline attributes $68.37\%$ of all version-based \ac{CVE} matches as \emph{false-positives} and filters them out of the result set.
In numbers - out of the median 1,196 matches per firmware, we reduce the set of potentially applicable \acp{CVE} to roughly 378.
Thus, we significantly reduce the result set of potentially applicable \acp{CVE} requiring manual verification by analysts and vendors.
The portion of low confidence applicability due to missing file references is $19.4\%$.
Unfortunately, for this class of matches our methodology does not generate added value.

\section{Limitations}
\label{sec:limitations}
We identify methodological shortcomings in three dimensions: applicability, sound ground truth, and functionality.

In terms of applicability, our Linux kernel \ac{CVE} attribution pipeline is bound to \ac{FACT}'s static analysis success.
If the kernel version, \ac{ISA}, and build configuration remain unknown, our method can not identify possibly included components for reliable \ac{CVE} filtering.
Then, the pipeline becomes inapplicable.
Yet, the case study in Section~\ref{sec:case_study} shows that (in the used firmware corpus) there is still a considerable amount of firmware fulfilling all requirements.

As for sound ground truth, reliable and true-positive \ac{CVE} attribution is limited by the quality of its underlying dataset.
If Linux kernel \ac{CVE} records do not reference truly affected versions and source files, the proposed mechanisms introduce false-positive, but also false-negative matches.
In this case, result reliability is negatively affected.
Our assumption of vendors using mainline kernels is another limiting factor that affects reliability, but a methodical necessity due to missing insider information.
Vendors may cherry-pick patches or introduce custom fixes, which are not detectable by our approach.
While some of the modifications might be obtainable through \ac{GPL} portals, we identify the issue of scalable accessibility.

Regarding functional limits, we stress the inherited limitations of static analyses.
They may use heuristics to find indicators of possible bug presence, but can hardly serve definitive proof -- which requires triggering the bug.

Finally, the conducted case study is limited in its validity, as the used corpus is missing in device class heterogeneity.

\section{Conclusion}
\label{sec:conclusion}
In this paper, we focused on improving result reliability of version-based \ac{CVE} matching for the special case of binary Linux kernels in large-scale static firmware analyses.
Heterogeneous hardware properties, modularity, numerous development streams, and vendor-specific builds cause high false-positive rates.
This is because the attribution method does not check for vulnerable component presence in binary images when filtering \acp{CVE}.

Despite general automation issues due to unsound or incomplete data in \ac{CVE} repositories and common challenges in binary firmware analyses, we found supplementary filter data in Linux issue descriptions and firmware samples to reduce the set of false-positive matches in scale.
We enriched naive version-based \ac{CVE} matching with a static attribution pipeline that detects kernel configurations and \acp{ISA} in firmware images to reconstruct the kernel build process and guess included source files.
This data then serves as filter using kernel \acp{CVE} where affected files are explicitly stated.

The case study shows that, with the limitations discussed in Section~\ref{sec:limitations} in mind, our approach is scalable and moderately applicable:
For $34.64\%$ of all 127 considered home router firmware images, the technical requirements are fulfilled and
about $65\%$ of all Linux kernel \acp{CVE} reference affected files in their description.

With naive version-based matching as baseline, the introduced approach generates a high confidence filter verdict for $80.6\%$ of all considered \acp{CVE} and reduces the result set by $68.37\%$:
\acp{CVE} affecting components probably absent from binary kernel samples are successfully eliminated.
While a non-negligible amount of \acp{CVE} that our method can not reliably filter remains, we conclude that the proposed attribution pipeline is a promising step towards more reliable and scalable static \ac{CVE} attribution.
Security analysts are still required to verify the true applicability of a bug, but our systematic and automated filtering may at least reduce their manual efforts.

Stage one of our pipeline is part of the publicly available~\ac{FACT}~\cite{fact}, stage two is a set of standalone scripts available at \url{https://github.com/fkie-cad/cve-attribution-s2}, and our case study corpus is documented for reconstruction at \url{https://github.com/fkie-cad/embedded-evaluation-corpus}.

\section{Future Work}

In future work, we want to address missing device class heterogeneity present in our case study corpus.
Aside of identifying and implementing additional filter criteria, we evaluate options to combine the proposed methodology with \emph{linuxkernelcves.org}~\cite{linuxkernelcves}.
The \ac{CPE} data used for version matching is of varying quality~\cite{unreliable_cpe} and may introduce false-positives and -negatives not caught by our methodology.
A more fine-granular commit-based version tracking as offered by~\cite{linuxkernelcves} is a promising addition to our pipeline.
However, a thorough review of the automatically applied method is required before adapting this data source over the de-facto standard of the NVD.

\bibliographystyle{IEEEtran}
\bibliography{ref}

\end{document}